\documentclass[aps,nofootinbib,preprintnumbers]{revtex4-2}
\usepackage{eurosym}
\usepackage{lipsum}
\usepackage{CJK}
\usepackage{amsfonts}
\usepackage{amsmath}
\usepackage{amssymb}
\usepackage[english]{babel}
\usepackage{graphicx}
\usepackage{epsfig}
\usepackage{bm}
\usepackage{verbatim}
\usepackage[utf8]{inputenc}
\usepackage{booktabs}
\usepackage{multirow}
\usepackage{subfig}
\usepackage{slashed}
\usepackage{xcolor}
\usepackage[colorlinks=true,urlcolor=red,citecolor=red]{hyperref}
\usepackage[font=small]{caption}
\usepackage{float}
\usepackage{blindtext}
\usepackage{placeins}
\usepackage[utf8]{inputenc}
\usepackage{bbm}
\usepackage[colorinlistoftodos]{todonotes}

\newenvironment{Eqnarray}{\arraycolsep 0.14em\begin{eqnarray}}{\end{eqnarray}}
\newcommand{\ba}{\begin{Eqnarray}}
\newcommand{\ea}{\end{Eqnarray}}
\newcommand{\be}{\begin{equation}}
\newcommand{\ee}{\end{equation}}
\newcommand{\bal}{\begin{aligned}}
\newcommand{\eal}{\end{aligned}}
\newcommand{\bea}{\begin{eqnarray}}
\newcommand{\eea}{\end{eqnarray}}
\newcommand{\ben}{\begin{enumerate}}
\newcommand{\een}{\end{enumerate}}
\newcommand{\bit}{\begin{itemize}}
\newcommand{\eit}{\end{itemize}}
\newcommand{\bde}{\begin{widetext}}
\newcommand{\ede}{\end{widetext}}

\renewcommand{\[}{\left[}

\def\lsim{\mathrel{\rlap{\lower4pt\hbox{\hskip1pt$\sim$}}
    \raise1pt\hbox{$<$}}}
\def\gsim{\mathrel{\rlap{\lower4pt\hbox{\hskip1pt$\sim$}}
    \raise1pt\hbox{$>$}}}

\def\3211{$\mathrm{SU(3) \otimes SU(2)_L \otimes U(1)_R \otimes U(1)_{B-L}}$ }
\def\321{$\mathrm{SU(3) \otimes SU(2) \otimes U(1)}$ }
\def\422{$\mathrm{SU(4) \otimes SU(2) \otimes SU(2)_R}$ }

\newcommand{\mathsym}[1]{{}}

\definecolor{bostonuniversityred}{rgb}{0.8, 0.0, 0.0}

\input{tcilatex}

\begin{document}

\title{Predictive linear seesaw model with $\Delta \left( 27\right) $ family symmetry.}
\author{A. E. C\'arcamo Hern\'andez$^{a,b,c}$}
\email{antonio.carcamo@usm.cl}
\affiliation{$^{{a}}$Universidad T\'ecnica Federico Santa Mar\'{\i}a, Casilla 110-V,
Valpara\'{\i}so, Chile\\
$^{{b}}$Centro Cient\'{\i}fico-Tecnol\'ogico de Valpara\'{\i}so, Casilla
110-V, Valpara\'{\i}so, Chile\\
$^{{c}}$Millennium Institute for Subatomic Physics at the High-Energy
Frontier, SAPHIR, Calle Fern\'andez Concha No 700, Santiago, Chile}
\author{Ivo de Medeiros Varzielas}
\email{ivo.de@udo.edu}
\affiliation{CFTP, Departamento de F\'{\i}sica, Instituto Superior T\'{e}cnico,
Universidade de Lisboa, Avenida Rovisco Pais 1, 1049 Lisboa, Portugal}
\author{Juan Marchant Gonz\'alez}
\email{juan.marchant@upla.cl}
\affiliation{Laboratorio de C\'omputo de F\'isica (LCF-UPLA), Facultad de Ciencias Naturales y Exactas, Universidad de Playa Ancha, Subida Leopoldo Carvallo 270, Valpara\'iso, Chile.}
\affiliation{Departamento de F\'{\i}sica, Universidad T\'ecnica Federico Santa Mar\'{\i}a%
\\
Casilla 110-V, Valpara\'{i}so, Chile}
\affiliation{Millennium Institute for Subatomic Physics at High-Energy Frontier (SAPHIR),
Fern\'andez Concha 700, Santiago, Chile}


\begin{abstract}
	We consider a model that accounts for the smallness of neutrino masses through the linear seesaw mechanism and employs a $\Delta(27)$ family symmetry to address the flavour problem. The model is predictive in the leptonic sector and faces constraints from Lepton Flavour Violation processes, namely $\mu \to e \gamma$, which indicate a range for the Right-Handed neutrino mass.
\end{abstract}

\maketitle



\section{Introduction}

\label{intro}

The observation of the Higgs boson at the LHC has further confirmed the Standard Model (SM). However some beyond SM physics is needed due to the observed neutrino mixing. Further, in the SM, the fermion masses and mixings arise from the Yukawa couplings of the fermions to the Higgs, which are numerous, hierarchical, and with a strikingly different pattern between quark mixing and the mixing observed for the leptons.
This is often referred to as the flavour problem, and a possible solution is that family symmetries are added, together with some type of seesaw mechanism. With the family symmetry, the Yukawa couplings to the SM Higgs are then obtained from the pattern of family symmetry breaking.

A popular family symmetry is $\Delta(27)$. Its main advantage is that it is a small symmetry with a triplet and anti-triplet representation. The typical pattern of symmetry breaking that emerges from $\Delta(27)$ is also able to naturally account for the observed leptonic mixing in a large class of models, and it also has very appealing interplay with CP symmetries.
\cite{Branco:1983tn, deMedeirosVarzielas:2006fc, Ma:2006ip, Ma:2007wu, Bazzocchi:2009qg, deMedeirosVarzielas:2011zw, Varzielas:2012nn, Bhattacharyya:2012pi, Ferreira:2012ri, Ma:2013xqa, Nishi:2013jqa, Varzielas:2013sla, Aranda:2013gga, Varzielas:2013eta, Harrison:2014jqa, Ma:2014eka, Abbas:2014ewa, Abbas:2015zna, Varzielas:2015aua, Bjorkeroth:2015uou, Chen:2015jta, Vien:2016tmh, Hernandez:2016eod,  Bjorkeroth:2016lzs, CarcamoHernandez:2017owh, deMedeirosVarzielas:2017sdv, Bernal:2017xat, CarcamoHernandez:2018iel, deMedeirosVarzielas:2018vab, CarcamoHernandez:2018hst, CarcamoHernandez:2018djj, Ma:2019iwj, Bjorkeroth:2019csz, CarcamoHernandez:2020udg}.

In this paper we consider a simple model that employs a $\Delta(27)$ family symmetry and linear seesaw mechanism to generate the tiny masses for the active neutrinos.

In order to obtain a viable model, we need to break $\Delta(27)$ with multiple fields \cite{Bhattacharyya:2012pi, Varzielas:2013sla, Varzielas:2013eta} (this is typical of discrete symmetries with triplets, see \cite{Felipe:2013vwa}).

This model of leptonic mixing we do not address the quarks, which can be fitted as in the SM.

The layout of this paper is as follows.  In Section \ref{sec:model} we describe the model's content and associated interactions. In Section \ref{sec:lmm} we present the lepton mass terms and respective mass matrices. In Section \ref{sec:lfv} we discuss constraints on the model. We conclude in Section \ref{sec:con}.

\section{The model \label{sec:model}}

We consider a supersymmetric model where the full symmetry $\mathcal{G}$
features a two-step spontaneous breaking pattern: 
\begin{eqnarray}
&&\mathcal{G}=SU(3)_{C}\times SU(2)_{L}\times U\left( 1\right) _{Y}\times
\Delta (27)\times Z_{N}\times Z_{9}  \notag \\
&&\hspace{35mm}\Downarrow \Lambda _{int}  \notag \\[3mm]
&&\hspace{15mm}SU\left( 3\right) _{C}\times SU\left( 2\right) _{L}\times\left( 1\right) _{Y}  \notag \\[3mm]
&&\hspace{35mm}\Downarrow v  \notag \\[3mm]
&&\hspace{23mm}SU\left( 3\right) _{C}\otimes U\left( 1\right) _{em}
\label{Group}
\end{eqnarray}%
It is assumed that the discrete groups are spontaneously broken at an energy
scale $\Lambda _{int}$ much larger than the electroweak symmetry breaking
scale $v=246$ GeV. In our supersymmetric model, the scalar sector of MSSM is
enlarged by several gauge singlet scalar fields, whereas the fermion sector
is augmented by two heavy RH Majorana neutrinos, i.e., $\nu _{R}$ and $N_{R}$. Such extension of the particle spectrum is necessary for the
implementation of the linear seesaw mechanism that yields the tiny active neutrino masses. The $\Delta (27)\times Z_{N}\times Z_{9}$ assignments of fermions and scalars in our model are shown in Table~\ref{tab:asig}. In our proposed model, the $Z_{N}$ symmetry allows that only one $\Delta (27)$ scalar triplet $\zeta $ participates in the charged lepton Yukawa interactions, thus yielding a diagonal charged lepton mass matrix, thanks to the vacuum expectation value (VEV) configuration of $\zeta $ to be shown below. Furthermore, $Z_{N}$ symmetry separates the $\Delta (27)$ scalar triplets $\rho $ and $\eta $ participating in the neutrino Yukawa interactions with $N_{R}$ with the $\Delta (27)$ triplet $\zeta $ appearing in the Yukawa terms for charged leptons and right handed
Majorana neutrino $\nu _{R}$. The $Z_{9}$ symmetry, which is spontaneously
broken by the VEV of the scalar singlet $\sigma $%
, is crucial for the implementation of the Froggat-Nielsen mechanism that
yields the observed patttern of SM charged lepton masses. Furthermore, the $%
\Delta (27)$ discrete flavor symmetry is necessary to get a predictive
lepton sector through the 
following VEV configurations for the $\Delta
\left( 27\right) $ triplets scalars: 
\begin{equation}
\left\langle \zeta \right\rangle =v_{\zeta }\left( 1,0,0\right) ,\hspace{1cm}%
\left\langle \rho \right\rangle =v_{\rho }\left( 1,0,-1\right) ,\hspace{1cm}%
\left\langle \eta \right\rangle =v_{\eta }\left( 0,1,0\right) ,
\label{D27VEVS}
\end{equation}%
where the VEV pattern of $\zeta $ yields a diagonal matrix for SM charged
leptons, thus implying that the leptonic mixing entirely arises from the
neutrino sector. These special VEV patterns are the motivation for
considering $\Delta (27)$ flavored SUSY models, as such configurations can
arise from the F-term alignment mechanism \cite{Varzielas:2015aua} or D-term
alignment mechanism \cite{deMedeirosVarzielas:2006fc}. This is in contrast
with the somewhat generic $(r,e^{i\psi },e^{-i\psi })$ VEV used in \cite%
{CarcamoHernandez:2017owh, CarcamoHernandez:2018hst}.

Since the observed pattern of SM charged lepton masses arises from the
spontaneous breaking of the $Z_{9}$ discrete symmetry, we set the VEVs of the different gauge singlet scalars as follows: 
\begin{equation}
v_{\rho }\sim v_{\sigma }\sim v_{\zeta }\sim v_{\eta }\sim \lambda \Lambda ,
\label{VEVhierarchy}
\end{equation}%
where $\lambda =\sin \theta _{13}$, being $\theta _{13}$ the reactor mixing
angle, whereas $\Lambda $ is the model cutoff, which can be interpreted as
the scale of the UV completion of the model, e.g. the masses of the
Froggatt-Nielsen messenger fields.

With the above mentioned particle content and symmetries, the following
leptonic Yukawa interactions arise: 
\begin{equation}
-\mathcal{L}_{Y}^{\left( l\right) }=y_{1}^{\left( l\right) }\left( \overline{%
l}_{L}H_{d}\zeta \right) _{\mathbf{1}_{0,0}}l_{1R}\frac{\sigma ^{8}}{\Lambda
^{9}}+y_{2}^{\left( l\right) }\left( \overline{l}_{L}H_{d}\zeta \right) _{%
\mathbf{1}_{0,2}}l_{2R}\frac{\sigma ^{4}}{\Lambda ^{5}}+y_{3}^{\left(
l\right) }\left( \overline{l}_{L}H_{d}\zeta \right) _{\mathbf{1}_{0,1}}l_{3R}%
\frac{\sigma ^{2}}{\Lambda ^{3}}+H.c. \label{Lyl}
\end{equation}%
\begin{eqnarray}
-\mathcal{L}_{Y}^{\left( \nu \right) } &=&y_{1}^{\left( \nu \right) }\left( 
\overline{l}_{L}H_{u}\rho \right) _{\mathbf{1}_{0,0}}\nu _{R}\frac{1}{%
\Lambda }+y_{2}^{\left( \nu \right) }\left( \overline{l}_{L}H_{u}\eta
\right) _{\mathbf{1}_{0,0}}\nu _{R}\frac{1}{\Lambda }  \notag \\
&&+y_{3}^{\left( \nu \right) }\left( \overline{l}_{L}H_{u}\zeta \right) _{%
\mathbf{1}_{0,0}}N_{R}\frac{\varphi _{1}}{\Lambda ^{2}}+\left( \overline{l}%
_{L}H_{u}\zeta \right) _{\mathbf{1}_{0,2}}N_{R}\frac{\varphi _{2}}{\Lambda
^{2}}+\left( \overline{l}_{L}H_{u}\zeta \right) _{\mathbf{1}_{0,3}}N_{R}%
\frac{\varphi _{3}}{\Lambda ^{2}}  \notag \\
&&+M\overline{\nu }_{R}N_{R}^{C}+H.c
\end{eqnarray}

\begin{table}
\centering
\begin{tabular}{c|cccccc|ccccccccc}
\hline\hline
& $l_{L}$ & $l_{1R}$ & $l_{2R}$ & $l_{3R}$ & $\nu _{R}$ & $N_{R}$ & $H_{u}$ & $H_{d}$ & 
$\sigma$ & $\zeta$ & $\rho$ & $\eta$ & $\varphi _{1}$ & $\varphi _{2}$ & $\varphi _{3}$ \\ 
\hline\hline
$\Delta \left( 27\right)$ & $\mathbf{3}$ & $\mathbf{1}_{0,0}$ & $\mathbf{1}_{0,1}$ & 
$\mathbf{1}_{0,2}$ & $\mathbf{1}_{0,0}$ & $\mathbf{1}_{0,0}$ & $\mathbf{1}_{0,0}$ & 
$\mathbf{1}_{0,0}$ & $\mathbf{1}_{0,0}$ & $\mathbf{3}$ & $\mathbf{3}$ & $\mathbf{3}$ & 
$\mathbf{1}_{0,0}$ & $\mathbf{1}_{0,1}$ & $\mathbf{1}_{0,2}$ \\ 
$Z_{N}$ & 0 & $-x$ & $-x$ & $-x$ & $2x$ & $-2x$ & 0 & 0 & 0 & x & $-2x$ & $-2x$ & $x$ & $x$ & $x$
\\ 
$Z_{9}$ & 0 & 8 & 4 & 2 & 0 & 0 & 0 & 0 & -1 & 0 & 0 & 0 & 0 & 0 & 0 \\
\hline\hline
\end{tabular}
\caption{Charge assignments of scalar and lepton fields}
\label{tab:asig}
\end{table}

\section{Lepton masses and mixings \label{sec:lmm}}
After the spontaneus breaking of the SM gauge symmetry and the $\Delta
(27)\times Z_{N}\times Z_{9}$ discrete group, we find that the SM charged
lepton mass matrix is diagonal with the charged lepton masses given by: 
\begin{equation}
m_{e}=y_{1}^{\left( l\right) }\frac{v_{\zeta }v_{\sigma }^{8}v_{h_{d}}}{%
\sqrt{2}\Lambda ^{9}}=a_{1}^{\left( l\right) }\lambda ^{9}\frac{v}{\sqrt{2}},%
\hspace{1cm}m_{\mu }=y_{2}^{\left( l\right) }\frac{v_{\zeta }v_{\sigma
}^{4}v_{h_{d}}}{\sqrt{2}\Lambda ^{5}}=a_{2}^{\left( l\right) }\lambda ^{5}%
\frac{v}{\sqrt{2}},\hspace{1cm}m_{\tau }=y_{3}^{\left( l\right) }\frac{%
v_{\zeta }v_{\sigma }^{2}v_{h_{d}}}{\sqrt{2}\Lambda ^{3}}=a_{3}^{\left(
l\right) }\lambda ^{3}\frac{v}{\sqrt{2}}\label{eq:lepmass}
\end{equation}
where $a_{1}^{\left( l\right) }$, $a_{2}^{\left( l\right) }$ and $%
a_{3}^{\left( l\right) }$ are real $\mathcal{O}(1)$ dimensionless parameters
and we have assumed that $v_{h_{d}}\sim v/\sqrt{2}$, being $v=246$ GeV the
electroweak symmetry breaking scale.

Furthermore, the neutrino Yukawa interactions yield the following neutrino mass terms: 
\begin{equation}
-\mathcal{L}_{mass}^{\left( \nu \right) }=\frac{1}{2}\left( 
\begin{array}{ccc}
\overline{\nu _{L}^{C}} & \overline{\nu _{R}} & \overline{N_{R}}%
\end{array}%
\right) M_{\nu }\left( 
\begin{array}{c}
\nu _{L} \\ 
\nu _{R}^{C} \\ 
N_{R}^{C}%
\end{array}%
\right) +H.c,
\end{equation}%
where the neutrino mass matrix is given by: 
\begin{equation}
M_{\nu }=\left( 
\begin{array}{ccc}
0_{3\times 3} & m_{1} & m_{2} \\ 
m_{1}^{T} & 0 & M \\ 
m_{2}^{T} & M & 0%
\end{array}%
\right) ,  \label{Mnu}
\end{equation}%
and the submatrices $m_{1}$ and $m_{2}$ have the following structure: 
\begin{equation}
m_{1}=\left( 
\begin{array}{c}
a \\ 
b \\ 
-a%
\end{array}%
\right) ,\hspace{1cm}m_{2}=\left( 
\begin{array}{c}
c \\ 
d \\ 
f%
\end{array}%
\right) .  \label{Mnublocks0}
\end{equation}%
The active light neutrino masses are generated from linear seesaw mechanism, and the physical low energy neutrino mass matrix is given by: 
\begin{eqnarray}
\widetilde{M}_{\nu }&=&-\left[ m_{1}M^{-1}m_{2}^{T}+m_{2}M^{-1}m_{1}^{T}\right], \notag\\
\widetilde{M}_{\nu }&=& \frac{1}{M}\left(\begin{array}{ccc}
2ac & ad + bc & a(f-c) \\
ad+bc & 2bd & bf-ad \\
a(f-c) & bf-ad & -2af 
\end{array}\right).  
\label{eq:neutrinamass}
\end{eqnarray}

Furthermore, the masses for the physical heavy pseudo-Dirac neutrinos are given by: 
\begin{eqnarray}
M_{N^{-}} &=&-M-m_{1}^{T}m_{1}M^{-1}+\frac{1}{2}\left[
m_{1}M^{-1}m_{2}^{T}+m_{2}M^{-1}m_{1}^{T}\right] , \\
M_{N^{+}} &=&M+m_{1}^{T}m_{1}M^{-1}+\frac{1}{2}\left[
m_{1}M^{-1}m_{2}^{T}+m_{2}M^{-1}m_{1}^{T}\right] ,
\end{eqnarray}

\begin{table}
\begin{tabular}{c|cccccc} 
\hline\hline
\text{Observable}      & \text{$\Delta m_{21}^2[10^{-5}\;\text{eV}]$} & \text{$\Delta m_{31}^2[10^{-3}\;\text{eV}]$} & \text{$\sin^2\theta_{12}/10^{-1}$} & \text{$\sin^2\theta_{23}/10^{-1}$} & \text{$\sin^2\theta_{13}/10^{-2}$} & \text{$\delta_{\text{CP}}/^{\circ}$}  \\ 
\hline\hline
\text{Experimental value $\pm1\sigma$} & $7.50_{-0.20}^{+0.22}$     & $2.55_{-0.03}^{+0.02}$     & $3.18\pm 0.16$    & $5.74\pm 0.14$    & $2.200_{-0.062}^{+0.069}$    & $194_{-22}^{+24}$                   \\
\text{Best-fit model} & $7.45$      & $2.52$                   & $3.25$                & $5.73$       & $2.29$      & $189.2$                           \\
\hline\hline
\end{tabular}
\caption{Neutrino oscillation parameters for normal order (NO) from global fits~\cite{deSalas:2020pgw}.}
\label{tab:neutrinos}
\end{table}

The mass matrix of light active neutrinos from Eq. \eqref{eq:neutrinamass} is consistent with the experimental data of neutrino oscillations, reproducing the neutrino mass squared difference, the mixing angles and the leptonic CP violating phase. The result of our fit can be seen in the table \ref{tab:neutrinos} where it is observed that the best-fit point of the model is within the experimental range of $1\sigma$ for each observable. This adjustment of the parameters that reproduce the neutrino sector was obtained by minimizing $\chi^2$ function, which is defined as,
\begin{equation}
\chi_{\nu}^{2}=\frac{\left( m_{21}^{\exp }-m_{21}^{th}\right) ^{2}}{\sigma
_{m_{21}}^{2}}+\frac{\left( m_{31}^{\exp }-m_{31}^{th}\right) ^{2}}{\sigma
_{m_{31}}^{2}}+\frac{\left( s_{\theta _{12}}^{\exp }-s_{\theta
_{12}}^{th}\right) ^{2}}{\sigma _{s_{12}}^{2}}+\frac{\left( s_{\theta
_{23}}^{\exp }-s_{\theta _{23}}^{th}\right) ^{2}}{\sigma _{s_{23}}^{2}}+%
\frac{\left( s_{\theta _{13}}^{\exp }-s_{\theta _{13}}^{th}\right) ^{2}}{%
\sigma _{s_{13}}^{2}}+\frac{\left( \delta _{CP}^{\exp }-\delta
_{CP}^{th}\right) ^{2}}{\sigma _{\delta }^{2}}\;,  \label{ec:funtion_error}
\end{equation}

where $m_{i1}$ is the difference of the square of the neutrino masses (with $i=2,3$), $s_{\theta _{jk}}$ is the sine function of the mixing angles (with $ j,k=1,2,3$) and $\delta _{CP}$ is the CP violation phase. The superscripts represent the experimental (\textquotedblleft exp\textquotedblright) and theoretical (\textquotedblleft th\textquotedblright) values and the $\sigma $ are the experimental errors. Therefore, the minimization of $\chi_{\nu}^2$ gives us the following value,
\begin{equation}
\chi_{\nu}^{2}=0.352.\label{eq:chi-nu}
\end{equation}
%

\begin{figure}[tbp]
\centering
\subfloat[]{
\includegraphics[scale=0.25]{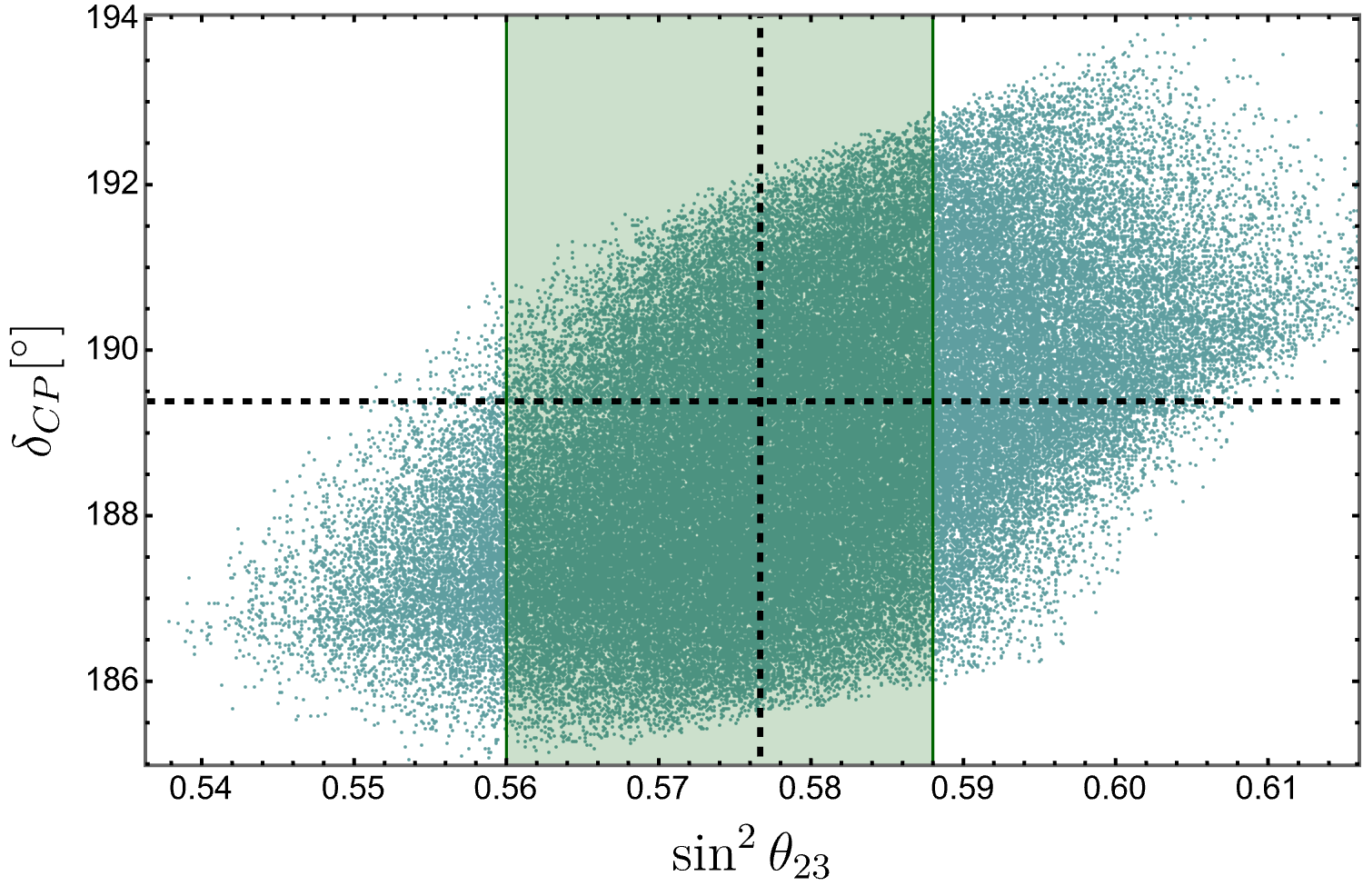}} \quad 
\subfloat[]{
\includegraphics[scale=0.25]{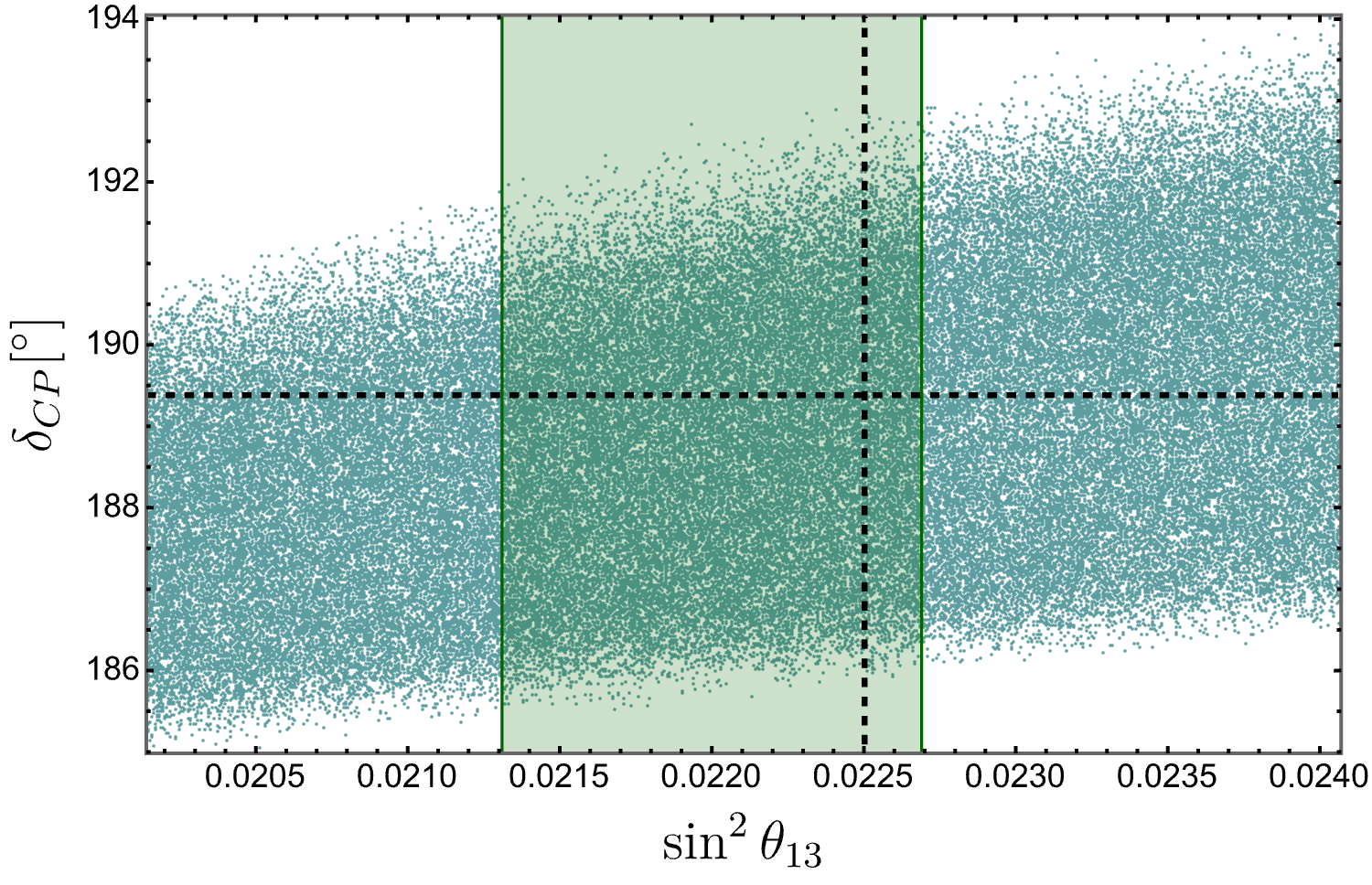}} \quad 
\subfloat[]{
\includegraphics[scale=0.25]{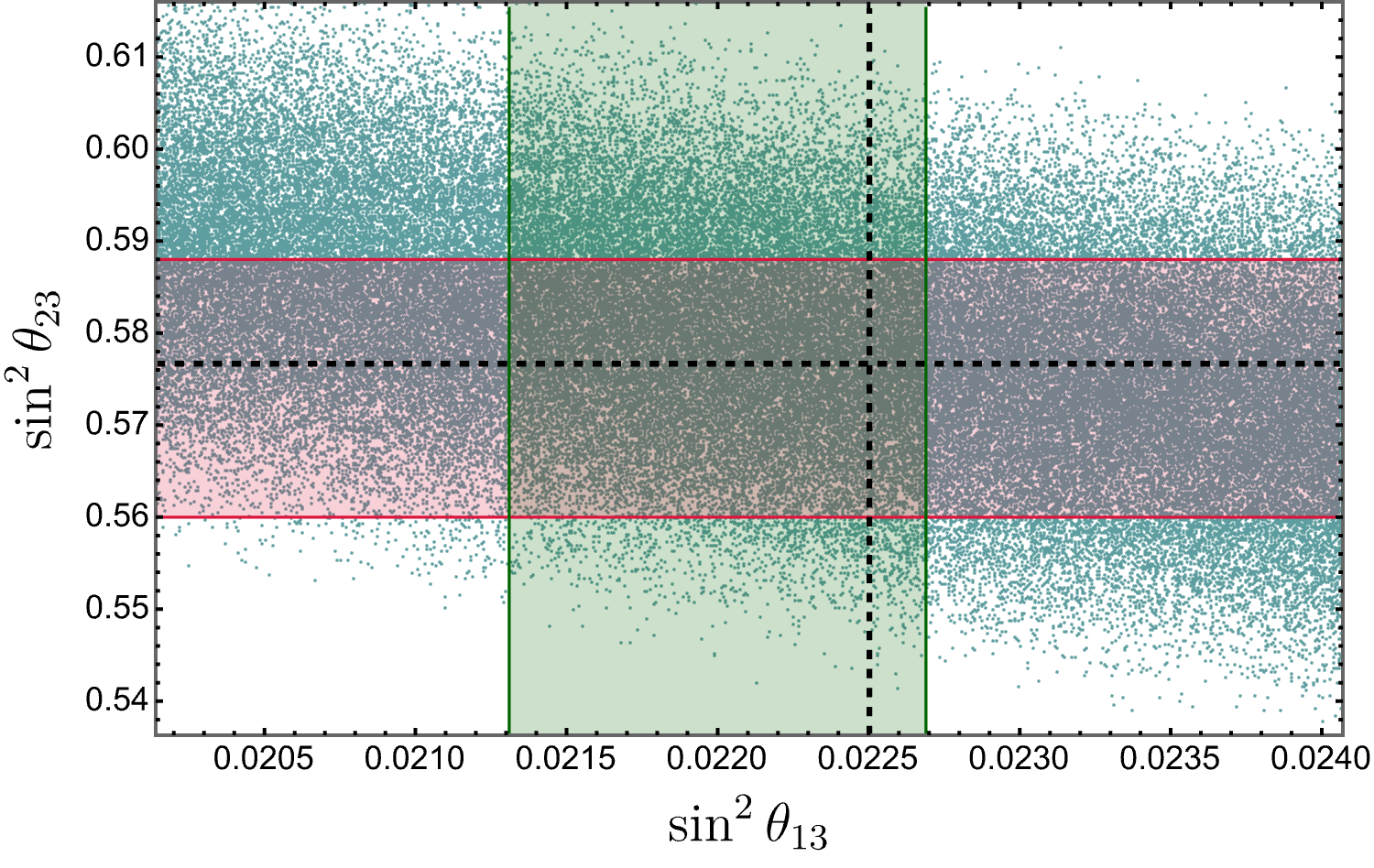}} \quad 
\subfloat[]{
\includegraphics[scale=0.25]{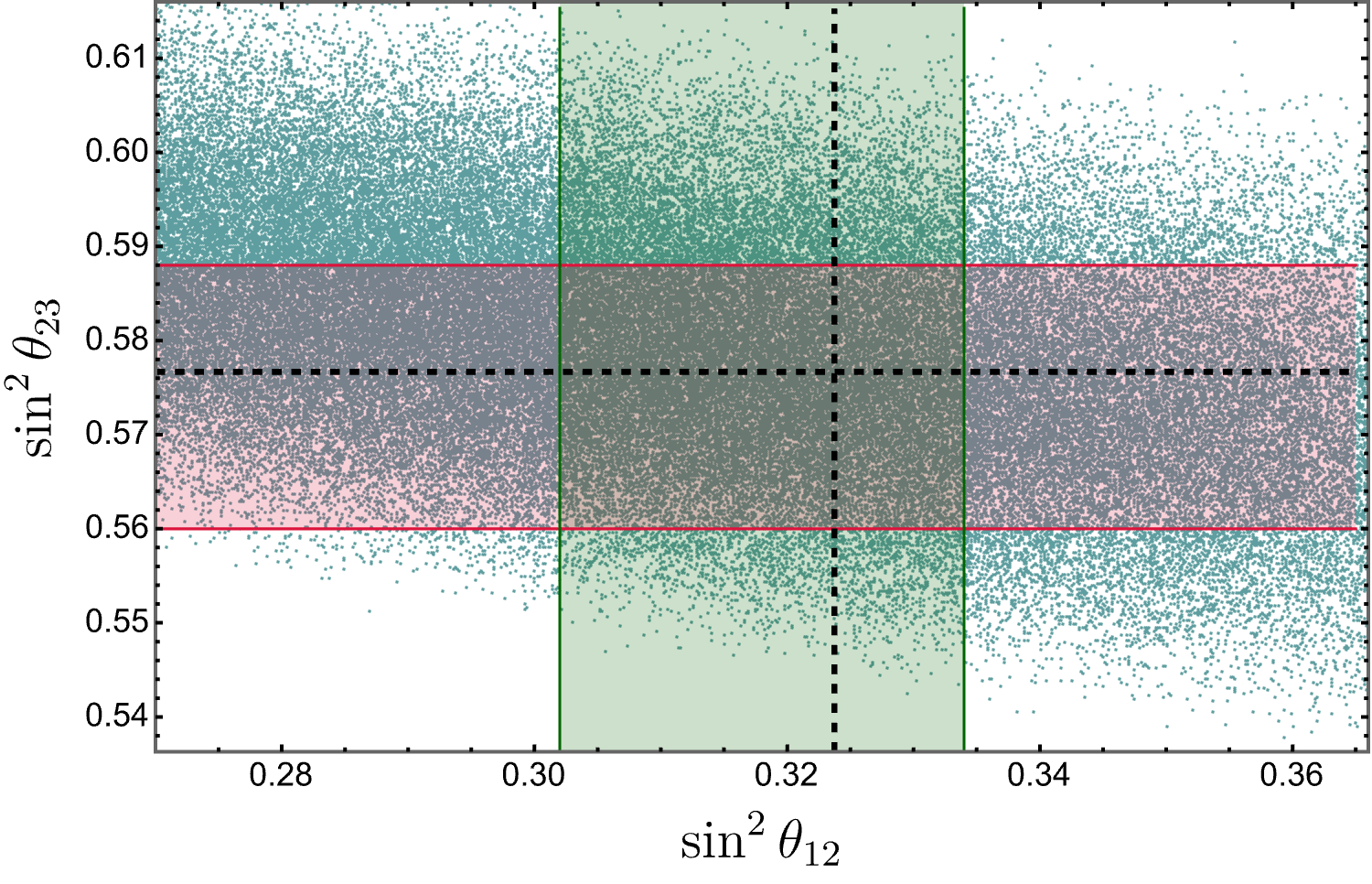}}
\caption{Correlation plot between $\sin^2\theta_{23}$, $\sin^2\theta_{13}$ with the leptonic Dirac CP violating phase (a and b, respectively) and $\sin^2\theta_{13}$, $\sin^2\theta_{12}$ with $\sin^2\theta_{23}$. The colored bands green and red correspond to the experimental range at $1\sigma$ and the black dotted lines represent the best-fit point value for each observable (c and d, respectively).}
\label{fig:corr-neutrino}
\end{figure}

On the other hand, by varying our parameters around 30\% of their best-fit point to generate data that are in the range of up to $3\sigma$, we can obtain correlations between the leptonic CP violating phase and the atmospheric (see Fig.~\ref{fig:corr-neutrino}a) and reactor mixing angles (Fig.~\ref{fig:corr-neutrino}b). In addition, we can also see a correlation between the solar mixing angle with the reactor and atmospheric mixing angles (Fig.~\ref{fig:corr-neutrino}c and \ref{fig:corr-neutrino}d, respectively), where our model obtains the following range of values for $0.536\lesssim \sin^2\theta_ {23}\lesssim 0.616$, $0.0201\lesssim \sin^2\theta_{13}\lesssim 0.0241$, $0.270\lesssim \sin^2\theta_{12}\leq 0.366$ and $185^{\circ}\lesssim \delta_{CP}\leq 194.1^{\circ}$, with all values of $\delta_{CP}$ being in the experimental range of $1\sigma$.

\begin{figure}
\centering
\includegraphics[scale=0.4]{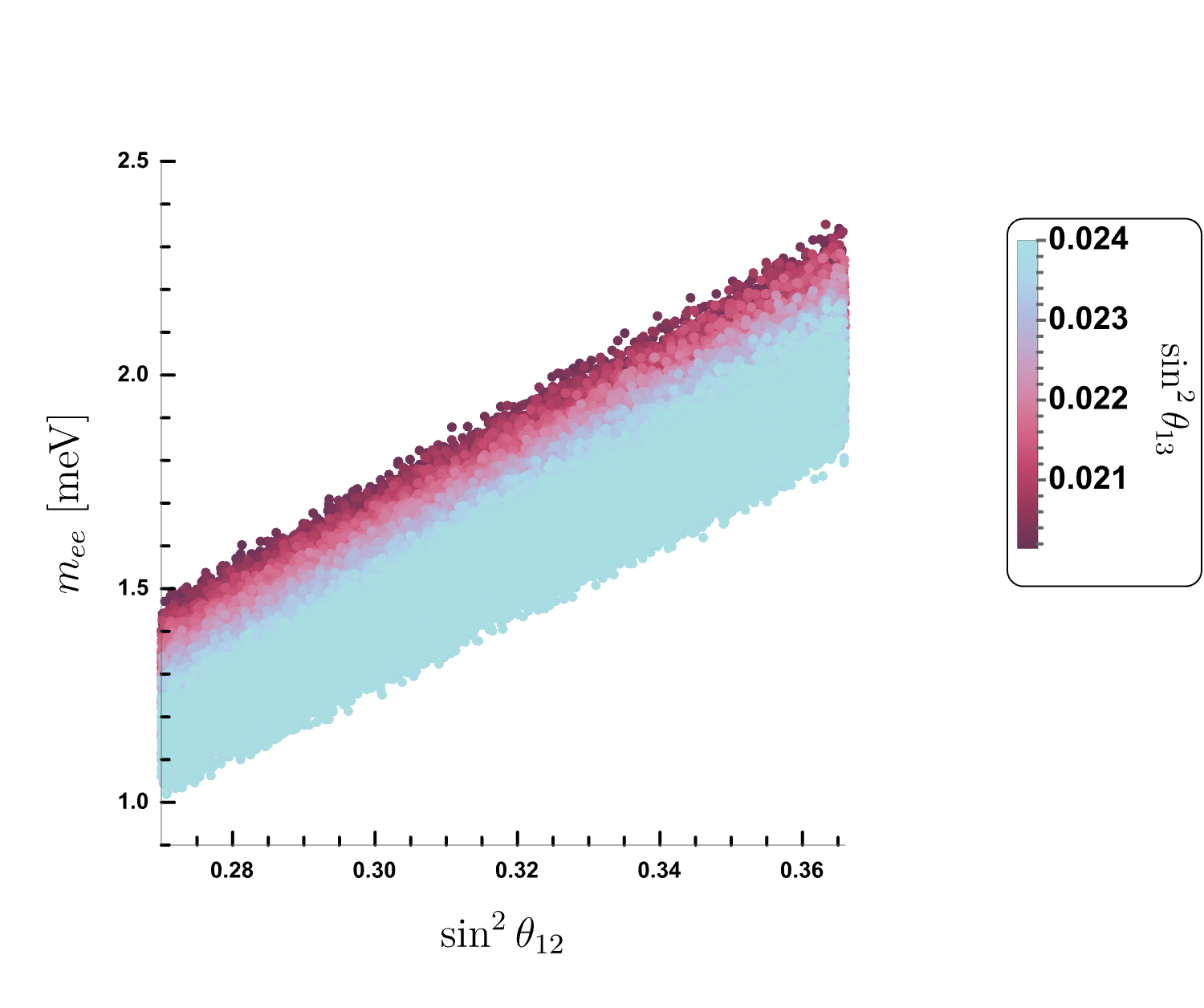}
\caption{Correlation between the solar mixing parameter $\sin^2\theta_{12}$ and the effective Majorana neutrino mass parameter $m_{ee}$ for different values of reactor mixing parameter.}
\label{fig:corr-mee}
\end{figure}

Furthermore, in our model we can obtain another observable, which corresponds to the effective Majorana neutrino mass parameter of the neutrinoless double beta decay without neutrinos. This observable provides information about the Majorana nature of neutrinos, where the shape of this mass parameter is given by:
\begin{equation}
m_{ee}=\left| \sum_i \mathbf{U}_{ei}^2m_{\nu i}\right|\;,
\label{ec:mee}
\end{equation}

where $\mathbf{U}_{ei}$ and $m_{\nu i}$ are the matrix elements of the PMNS leptonic mixing matrix and the light active neutrino masses, respectively. The neutrinoless double beta ($0\nu\beta\beta$) decay amplitude is proportional to $m_{ee}$. Fig.~\ref{fig:corr-mee} shows the correlation between the Majorana neutrino effective mass parameter $m_{ee}$ and the solar mixing parameter $\sin\theta_{12}$ considering different values of the reactor mixing parameter $\sin\theta_{13}$. As can be seen in Fig.~\ref{fig:corr-mee}, the model predicts an effective mass parameter of the Majorana neutrino in the range $1.02\; \text{meV}\lesssim m_{ee}\lesssim 2.35\; \text{meV}$. The current strictest experimental upper limit on the Majorana neutrino effective mass parameter, i.e., $m_{ee}\leq 50\; meV$ arises from the KamLAND-Zen boundary at $^{136}X_e\; 0\nu\beta\beta$ decay half-life $T_{1/2}^{0\nu\beta\beta}(^{136}X_e) >2.0\times 10^{26}$ year~\cite{KamLAND-Zen:2022tow}.

\section{Charged Lepton Flavor Violation \label{sec:lfv}}
In the next section, we discuss the charged lepton flavor violation (cLFV) processes present due to the mixing between active and heavy sterile neutrinos. In this analysis, we focus in the one-loop decay $l_i\rightarrow l_f\gamma$ where the branching ratios are given  by~\cite{Langacker:1988up, Lavoura:2003xp, Hue:2017lak}
\begin{eqnarray}
\text{BR}\left( l_{i}\rightarrow l_{j}\gamma \right) &=&\frac{\alpha
_{W}^{3}s_{W}^{2}m_{l_{i}}^{5}}{256\pi ^{2}m_{W}^{4}\Gamma _{i}}\left\vert
G_{ij}\right\vert ^{2}  \label{Brmutoegamma1} \\
G_{ij} &\simeq &\sum_{k=1}^{3}\left( \left[ \left( 1-RR^{\dagger }\right)
U_{\nu }\right] ^{\ast }\right) _{ik}\left( \left( 1-RR^{\dagger }\right)
U_{\nu }\right) _{jk}G_{\gamma }\left( \frac{m_{\nu _{k}}^{2}}{m_{W}^{2}}%
\right) +2\sum_{l=1}^{2}\left(R^{\ast }\right) _{il}\left( R\right)
_{jl}G_{\gamma }\left( \frac{m_{N_{R_l}}^{2}}{m_{W}^{2}}\right),
\label{Brmutoegamma2} \\
G_{\gamma } (x) &=&\frac{10-43x+78x^{2}-49x^{3}+18x^{3}\ln x+4x^{4}}{%
12\left( 1-x\right) ^{4}},  \notag
\end{eqnarray}
where $\Gamma _{\mu }=3\times 10^{-19}$ GeV is the total muon decay width, $U_{\nu}$ is the matrix that diagonalizes the light neutrinos mass matrix which, in our case, is equal to the PMNS matrix since the charged lepton mixing matrix is equal to the identity $U_{\ell}=\mathbb{I}$, as can be seen in Eq.~\eqref{eq:lepmass}. In addition, the matrix $R$ is given by 
\begin{equation}
R=\frac{1}{\sqrt{2}}m_1^{*}M^{-1},  \label{eq:Rneutrino}
\end{equation}

where $M$ and $m_1$ are the heavy Majorana mass and the Dirac neutrino submatrix, respectively.

The charged lepton flavor-violating processes $\tau\rightarrow \mu\gamma$ and $\tau\rightarrow e\gamma$ are highly suppressed in our model; however, for the process $\mu\rightarrow e\gamma$ the model predicts values below the experimental limit, its value for our best-fit point being equal to,
\begin{equation}
\text{BR}(\mu\rightarrow e\gamma)\simeq 4.11\times 10^{-13}.\label{eq:bestBR}
\end{equation}

\begin{figure}[h]
\centering
\includegraphics[scale=0.4]{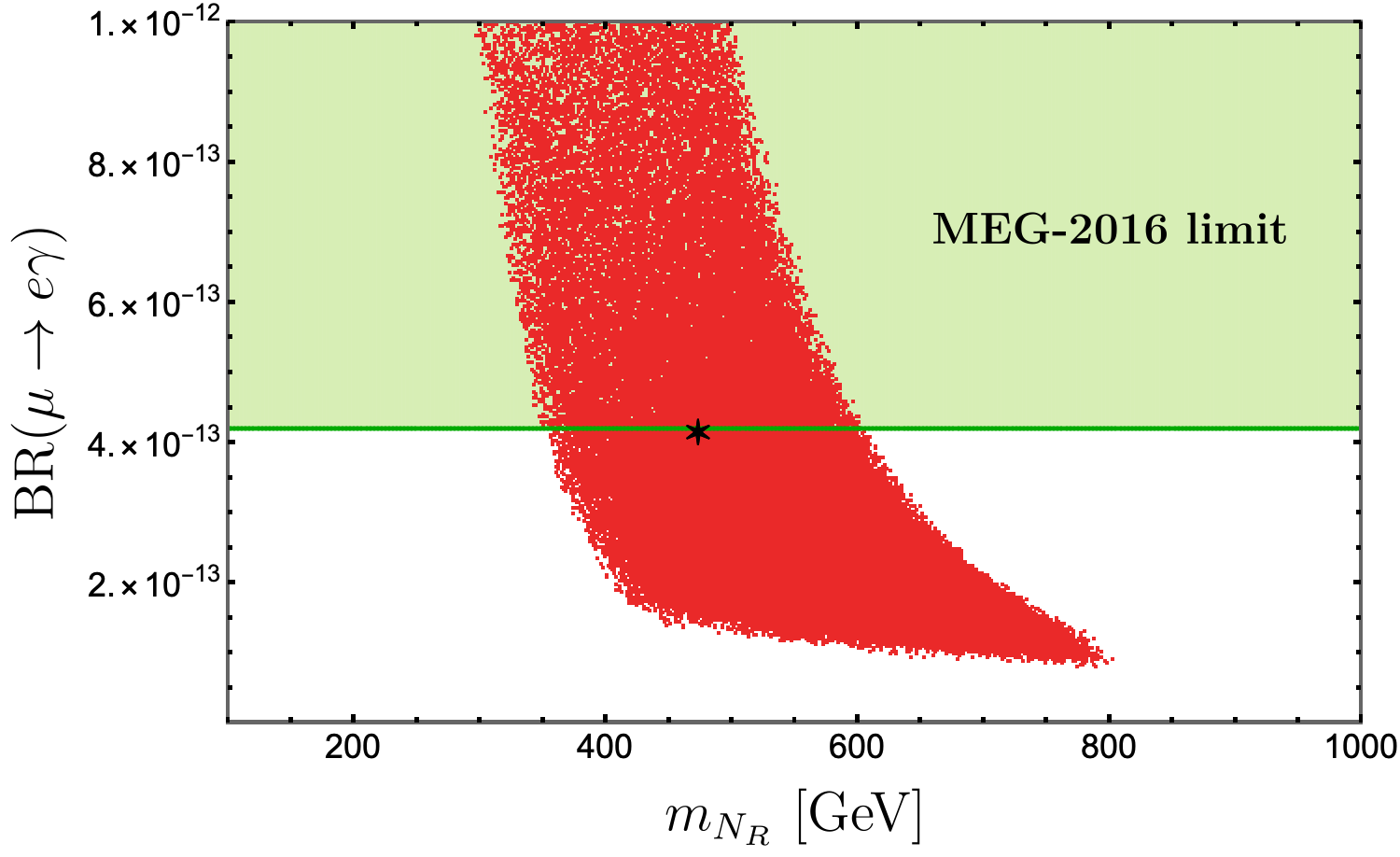}
\caption{Branching ratio for the process $\mu \rightarrow e\gamma$ as a function of the mass of the heavy pseudo-Dirac neutrino $N_R$. The shaded region corresponds to the values excluded by MEG~\cite{MEG:2016leq}. The red points are compatible with neutrino oscillation at $3\sigma$, while the black star corresponds to the model prediction for the best-fit point values.}
\label{fig:brmuegamma}
\end{figure}

Fig.~\ref{fig:brmuegamma} shows the correlation between the branching ratio $\text{BR}(\mu\rightarrow e\gamma)$ and the mass of the heavy pseudo-Dirac neutrino $N_R$. The green horizontal line and the shaded region in this plot represent the experimental limit given by MEG~\cite{MEG:2016leq} collaboration,
\begin{equation}
\text{BR}\left( \mu \rightarrow e\gamma \right)^{\text{exp}}<4.2\times 10^{-13}.
\end{equation}

The red scatter points in Fig.~\ref{fig:brmuegamma} represent the random variation of the parameters around 30\% of the best-fit point, while the black star corresponds to the value of the branching ratio given in Eq.~\eqref{eq:bestBR}. All red points are consistent with the neutrino oscillation experimental data at $3\sigma$, and set the mass of the heavy pseudo-Dirac neutrino pair in the range $236.7\;\text{GeV}\lesssim m_{ N_R}\lesssim 802.1 \; \text{GeV}$, where the lowest bound of $236.7$ \text{GeV} arises from requiring a $\mu\to e\gamma$ decay rate below its upper experimental limit.

\section{Conclusions \label{sec:con}}

We have proposed a supersymmetric model based on the $\Delta(27)$ family symmetry, supplemented by auxiliary cyclic symmetries where the tiny neutrino masses are generated from a linear seesaw mechanism. In the proposed theory the Standard Model charged lepton mass hierarchy arises from the spontaneous breaking of the cyclic symmetries. The $\Delta(27)$ family symmetry yields a diagonal charged lepton mass matrix in the symmetry basis, where the leptonic mixings entirely arises from the neutrino sector. Our model provides a successful fit of the experimental values for the neutrino mass squared splittings, leptonic mixing angles and leptonic Dirac CP phase. We have analyzed its consequences on charged lepton flavor violation and found that the considered model predicts very small rates for the $\tau\to\mu\gamma$ and $\tau\to e\gamma$ decays but sizeable rates for the $\mu\to e\gamma$ decay, within the reach of sensitivity of the forthcoming experiments, constraining the right-handed neutrino mass to be in a range.

\section*{Acknowledgments}

A.E.C.H is supported by ANID-Chile FONDECYT 1210378, ANID PIA/APOYO AFB230003, ANID- Programa Milenio - code ICN2019\_044, PIIC program of Universidad Técnica Federico Santa María and ANID Programa de Becas Doctorado Nacional code 21212041. IdMV acknowledges
funding from Funda\c{c}\~{a}o para a Ci\^{e}ncia e a Tecnologia (FCT)
through the contract UID/FIS/00777/2020 and was supported in part by FCT
through projects CFTP-FCT Unit 777 (UID/FIS/00777/2019),
PTDC/FIS-PAR/29436/2017, CERN/FIS-PAR/0004/2019 and CERN/FIS-PAR/0008/2019
which are partially funded through POCTI (FEDER), COMPETE, QREN and EU. AECH
thanks the Instituto Superior T\'{e}cnico, Universidade de Lisboa for
hospitality, where part of this work was done.

\appendix


\section{The $\Delta (27)$ discrete group}

\label{delta27} The $\Delta (27)$ discrete group has the following 11
irreducible representations: one triplet $\mathbf{3}$, one antitriplet $%
\overline{\mathbf{3}}$ and nine singlets $\mathbf{1}_{k,l}$ ($k,l=0,1,2$),
where $k$ and $l$ identify how the singlets transform under order 3
generators, corresponding to a $Z_{3}$ and $Z_{3}^{\prime }$ subgroups of $%
\Delta (27)$. 
\begin{eqnarray}
\mathbf{3}\otimes \mathbf{3} &=&\overline{\mathbf{3}}_{S_{1}}\oplus 
\overline{\mathbf{3}}_{S_{2}}\oplus \overline{\mathbf{3}}_{A}  \notag \\
\overline{\mathbf{3}}\otimes \overline{\mathbf{3}} &=&\mathbf{3}%
_{S_{1}}\otimes \mathbf{3}_{S_{2}}\oplus \mathbf{3}_{A}  \notag \\
\mathbf{3}\otimes \overline{\mathbf{3}} &=&\sum_{r=0}^{2}\mathbf{1}%
_{r,0}\oplus \sum_{r=0}^{2}\mathbf{1}_{r,1}\oplus \sum_{r=0}^{2}\mathbf{1}%
_{r,2}  \notag \\
\mathbf{1}_{k,\ell }\otimes \mathbf{1}_{k^{\prime },\ell ^{\prime }} &=&%
\mathbf{1}_{k+k^{\prime }mod3,\ell +\ell ^{\prime }mod3} \\
&&
\end{eqnarray}%
Denoting $\left( x_{1},y_{1},z_{1}\right) $ and $\left(
x_{2},y_{2},z_{2}\right) $ as the basis vectors for two $\Delta (27)$%
-triplets $\mathbf{3}$, one finds: 
\begin{eqnarray}
\left( \mathbf{3}\otimes \mathbf{3}\right) _{\overline{\mathbf{3}}_{S_{1}}}
&=&\left( x_{1}y_{1},x_{2}y_{2},x_{3}y_{3}\right) ,  \notag
\label{triplet-vectors} \\
\left( \mathbf{3}\otimes \mathbf{3}\right) _{\overline{\mathbf{3}}_{S_{2}}}
&=&\frac{1}{2}\left(
x_{2}y_{3}+x_{3}y_{2},x_{3}y_{1}+x_{1}y_{3},x_{1}y_{2}+x_{2}y_{1}\right) , 
\notag \\
\left( \mathbf{3}\otimes \mathbf{3}\right) _{\overline{\mathbf{3}}_{A}} &=&%
\frac{1}{2}\left(
x_{2}y_{3}-x_{3}y_{2},x_{3}y_{1}-x_{1}y_{3},x_{1}y_{2}-x_{2}y_{1}\right) , 
\notag \\
\left( \mathbf{3}\otimes \overline{\mathbf{3}}\right) _{\mathbf{1}_{r,0}}
&=&x_{1}y_{1}+\omega ^{2r}x_{2}y_{2}+\omega ^{r}x_{3}y_{3},  \notag \\
\left( \mathbf{3}\otimes \overline{\mathbf{3}}\right) _{\mathbf{1}_{r,1}}
&=&x_{1}y_{2}+\omega ^{2r}x_{2}y_{3}+\omega ^{r}x_{3}y_{1},  \notag \\
\left( \mathbf{3}\otimes \overline{\mathbf{3}}\right) _{\mathbf{1}_{r,2}}
&=&x_{1}y_{3}+\omega ^{2r}x_{2}y_{1}+\omega ^{r}x_{3}y_{2},
\end{eqnarray}%
where $r=0,1,2$ and $\omega =e^{i\frac{2\pi }{3}}$.

\bibliographystyle{utphys}
\bibliography{LSDelta27.bib}

\end{document}